\documentclass{article}
\usepackage{spconf,amsmath,graphicx}

\usepackage{enumitem}
\setlist{nosep, leftmargin=14pt}

\usepackage{caption}
\usepackage[font=small,skip=5pt]{caption}
\title{Multi-compartment diffusion MRI, T2 relaxometry and myelin water imaging as neuroimaging descriptors for anomalous tissue detection}
\name{Elda Fischi-Gomez$^{\star}$ \qquad Jonathan Rafael-Patino$^{\star}$ \qquad Marco Pizzolato$^{\diamond,\star}$ \qquad Gian Franco Piredda $^{\dagger, \star, \ddagger}$ \thanks{EFG and JRP equally contributed to this work.}}
\secondlinename{Tom Hilbert $^{\dagger}$ \qquad Tobias Kober $^{\dagger}$ \qquad Erick J. Canales-Rodriguez $^{\star}$ \qquad Jean-Philippe Thiran $^{\star, \ddagger}$ }
\address{\small$^{\star}$ Signal Processing Laboratory 5, Swiss Federal Institute of Technology Lausanne, Lausanne, Switzerland \\
    \small$^{\diamond}$Department of Applied Mathematics and Computer Science, Technical University of Denmark, Kongens Lyngby, Denmark\\
    \small$^{\dagger}$ Advanced Clinical Imaging Technology, Siemens Healthcare AG, Lausanne, Switzerland \\
    \small$^{\ddagger}$ Department of Radiology, Lausanne University Hospital and University of Lausanne, Switzerland }
\begin{document}
\maketitle
\vspace{-30pt}
\begin{abstract}
Multiple sclerosis (MS) is an inflammatory and neurodegenerative disease characterized by diffuse and focal areas of tissue loss. Conventional MRI techniques such as T1-weighted and T2-weighted scans are generally used in the diagnosis and prognosis of the disease. Yet, these methods are limited by the lack of specificity between lesions, its perilesional area and non-lesional tissue. Alternative MRI techniques exhibit a higher level of sensitivity to focal and diffuse MS pathology than conventional MRI acquisitions. However, they still suffer from limited specificity when considered alone. In this work, we have combined tissue microstructure information derived from multicompartment diffusion MRI and T2 relaxometry models to explore the voxel-based prediction power of a machine learning model in a cohort of MS patients and healthy controls. Our results show that the combination of multi-modal features, together with a boosting enhanced decision-tree based classifier, which combines a set of weak classifiers to form a strong classifier via a voting mechanism, is able to utilise the complementary information for the classification of abnormal tissue.
\end{abstract}

\vspace{-5.5mm}
\section{Introduction}
\vspace{-3.5mm}

\label{sec:intro}
Multiple Sclerosis (MS) is a chronic disease of the central nervous system (CNS) characterized by focal and diffuse inflammation and degeneration within the cerebral tissue \cite{Noseworthy2000}. Brain inflammation leads to myelin/axonal damage and altered axonal organization, which translates to the presence of diffuse demyelination, gliosis and axonal damage in non lesional normal-appearing (NA) tissue \cite{Noseworthy2000}. The pathological hallmark of MS is multiple focal areas of myelin loss, also known as plaques or lesions \cite{Kerschensteiner2004}.
MS has an intrinsic disseminated nature \cite{Kerschensteiner2004}, with individual patients differing from each other regarding clinical presentation, level of disability and number and anatomical locations of the demyelinating lesions. For MS assessment, MRI is the reference neuroimaging modality. However, the processes linked to MS pathology are difficult to visualize using conventional in-vivo imaging techniques, as they are generally limited by low pathological specificity and low sensitivity to diffused damage \cite{Cercigani2018,Erzinger2015,FilippiAgosta}.
Advanced quantitative MRI (qMRI) techniques provide complementary information about the different components of brain tissue architecture, and are potentially more sensitive and specific to different brain tissue abnormalities. In recent years, multiple studies have demonstrated the value of using multi-modal techniques. Examples of these are diffusion tensor imaging (DTI) and myelin water imaging \cite{Ilona2018}, magnetization transfer, DTI, T1 and T2 images \cite{Klistorner2018,Chatterjee2018, Bonnier2015}, and susceptibility-weighted imaging \cite{Haacke2009}, among others.
In this study, we propose to combine metrics derived from multi-component T2 relaxometry \cite{Mackay1994}, including the water fractions of the intra- and extra-cellular space, free water, and myelin compartments, with metrics derived from the Spherical Mean Technique (SMT) \cite{Kaden16}, an advanced multi-shell diffusion MRI (dMRI) model that decouples the signal fractions of the intra- and extra-axonal compartments.
This choice is motivated by the complementary nature of the information provided by these two techniques: whilst SMT enables to disentangle the intra- and extra-axonal compartments of the white matter \cite{Ferizi2015}, the contribution from the myelin compartment to the acquired signal is virtually zero; conversely, T2 relaxometry analysis does not allow to unambiguously separate the intra- and extra-axonal compartments, but it is sensitive to the myelin water \cite{Mackay1994}.
The goal of this study is two-fold: first, to implement a state-of-the-art machine learning method known for improving the prediction quality with small sample sizes (in MS lesions can be very sparse and data difficult to collect) to test whether such a handcrafted-based model can predict whole brain voxel-based lesions on multi-modal MRI datasets acquired without using external contrast agents; and second, to evaluate the prediction power of using multiple microstructure metrics derived from multi-shell dMRI and T2 relaxometry for the automatic classification of anomalous tissue.
\vspace{-4.5mm}
\section{Materials and Methods}
\vspace{-4mm}
\subsection{Data}
\vspace{-1.5 mm}
Twenty  early stage relapsing-remitting MS (RRMS) patients and twenty sex and age-matched healthy controls (HC) were scanned using the same 3T MRI Siemens scanner with a standard 32-channel head/neck coil. High-resolution human brain multi-echo T2 (MET2) data were collected using a prototype 3D multi-echo gradient and spin-echo sequence accelerated with CAIPIRINHA \cite{Piredda-ISMRM-2019} using the following parameters: matrix-size=144x126; voxel-size=1.8x1.8x1.8mm$^3$; $\Delta TE$/N-echoes/TR =10.68ms/32/1s; prescribed FA= 180$^0$; number-of-slices=84; acceleration factor=3x2; number of averages=1; acquisition time=10:30min.

The dMRI acquisition consisted in a four-shell protocol, including 6 images at $b$=700s/mm$^2$, 20 at $b$=1000s/mm$^2$, 46 at $b$=2000s/mm$^2$ and 66 at $b$=3000s/mm$^2$, and 11 intersperse $b=0$ ($b0$) images. The voxel-size was 1.8x1.8x1.8mm$^3$, and TE and TR were 75ms and 4500ms, respectively. Additionally, 12 reverse encoding $b0$ images were acquired for distortion correction. Standard MP2RAGE and FLAIR structural 3D images with voxel-size = 1x1x1mm$^3$ were also acquired for the initial radiologic evaluation of the subjects, and to segmented brain lesions \cite{LaRosa}. Susceptibility-induced distortions were eliminated using the $b0$ images acquired with reversed phase-encoding polarities, and the dMRI data were corrected for eddy-currents and subjects motion using FSL tools \cite{AnderssonSotiropoulos}.
The protocol was approved by the institutional review board, and all participants gave their written consent.
\vspace{-2mm}
\subsection{Multi-compartment microscopic diffusion MRI model}
\vspace{-1.5 mm}
The data were analysed by using an in-house python implementation of the Spherical Mean Technique (SMT) proposed by Kaden et al. \cite{Kaden16}. It models brain tissue per voxel into an intra-neurite domain that includes dendrites and axons, and the extra-neurite compartment that includes neurons, glial cells, and extracellular space. This technique provides estimates of neurite density unconfounded by fibre crossings and orientation dispersion. The features extracted from this model are the intra-neurite signal fraction $SMT_{f_{I}}$ and the extra-cellular signal fraction $SMT_{F_{E}}$.
\vspace{-2mm}
\subsection{Multi-component T2 reconstruction}
\vspace{-1.5 mm}
The Extended-Phase-Graph (EPG) model \cite{Prasloski2012_1} was employed to correct for stimulated echoes due to non-ideal experimental conditions. It allowed quantifying the deviation of the flip angle (FA) from the prescribed value, due to B1 field inhomogeneities and imperfect excitation pulses \cite{Hennig2004}. In our study, the dictionary matrix was built using the EPG model with $p=60$ T2-logarithmically-spaced points ranging from 10-2000 ms \cite{Prasloski2012_2}.

In a first step, the noise was partially suppressed by filtering the raw data with a 3D total variation algorithm \cite{chambolle2004}. The filtered data were corrected for head motion by linearly registering all image volumes with different echo times to the first volume. In a second step, different dictionary matrices were generated, each one corresponding to a fixed FA value from a discrete set between $90^0$ and $180^0$ equally spaced by $1^0$. The fitting process was carried out independently for each dictionary matrix by using the standard non-negative least squares (NNLS) method and the actual FA was determined as the one minimizing the mean squared error \cite{Prasloski2012_1}. Finally, the intra-voxel T2 distribution was calculated by using a regularized NNLS method \cite{Mackay1994,Whittall1997} based on a second-difference Laplacian matrix that promotes smooth solutions, as described in \cite{Erick-ISMRM-2019}. The optimal regularization parameter was determined by using the L-curve method \cite{Hansen1992}, as implemented in \cite{Castellanos2002}. The features extracted from the multi-compartment T2 reconstruction where the myelin water fraction ($MET2_{f_{M}}$), the intra- and extra-cellular water fraction combined in a single scalar ($MET2_{f_{IE}}$) and the free water fraction corresponding primarily to free fluids such as the cerebro-spinal fluid (CSF) ($MET2_{f_{csf}}$). The resulting images were registered to the diffusion MRI native space by using the acquired $b0$ images as reference. The non-linear registration was carried out with the Elastix software \cite{elastix}.
 \vspace{-2mm}
\subsection{Lesion identification}
\vspace{-1.5 mm}

\begin{figure}[htp]
    \centering
    \includegraphics[width=4.5cm]{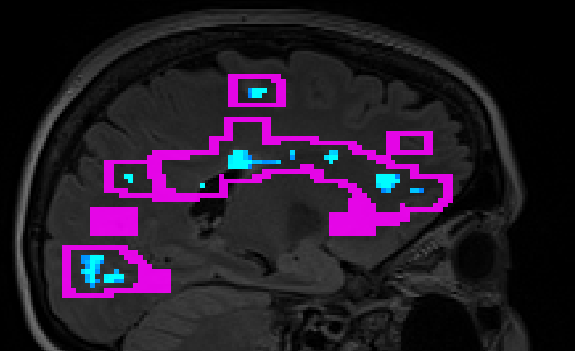}
    \caption{Extracted training data for the lesions (light blue) and NAWM voxels (purple). All voxels with a lesion score above $0.75$ where chosen as training data; NAWM voxels (N) where extracted from the surrounding WM of a neighborhood with minimum distance of six voxels from a lesion voxel (lesion score $> 0$). }
    \label{fig:lesion_voxels}
\end{figure}

Lesions were segmented on T1-w and T2-w images, and verified and rated by two trained neurologist. Specifically, the lesion masks consisted on voxel-wise probability distribution map, where "healthy" voxels have null probability of being a lesion \cite{LaRosa}. On the training set, lesional voxels (L) were selected as voxels having a lesion score higher than $0.75$ on the segmentation mask. This threshold was set to reduce the uncertainty in the lesion contours. A sharp limit was enforced in order to exclude perilesional voxels that may belong to normal appearing tissue. From the patients data, the initial set of MS lesions counting for a total number of 18,441 voxels were identified and later refined and randomly sampled to build the training dataset. The voxels considered as normal appearing WM (NAWM) were selected by defining an external surrounding ring to each MS lesion. NAWM voxels were defined as being at 6 voxels away for any lesion and named as (N) (Fig. \ref{fig:lesion_voxels}). Voxels considered as healthy tissue voxels (C) were selected from the control dataset.
 \vspace{-3.5mm}

\subsection{Identifying abnormal tissue voxels}
\vspace{-1.5 mm}
We performed voxel identification using the SMT-dMRI and MET2 estimates. From the patients data, we first left aside two subjects for validation which voxels where not included during the training sampling procedure. In the reminding database, we had 18,441 (L), 172,559 (N) and 103,614 (C) voxels, respectively. We employed a random shuffle and repeat strategy to compensate for the imbalanced number of voxels in each class. Finally, only lesion voxels with a score above $0.75$ where chosen to avoid mislabeled samples, at the cost of loosing training data. A total 8,455 lesions voxels remained after the pruning from which we used 5,919 (70\%) for training, while the remaining 2536 (30\%) voxels were used to evaluate the classifier performance. This method was repeated $k=5$ times in a k-fold validation strategy to estimate the model accuracy.
\vspace{-3.5mm}

\subsection{Machine learning based classification scores}
\vspace{-1.5 mm}

Boosting is a successful strategy used for assembling learning since first introduced by Freund \& Schapire~\cite{Boosting} in the Adaboost classifier; it combines a set of weak classifiers to form a strong classifier via an aggregation-based voting mechanism, which in addition results in a probability of classification estimate. The first weak classifiers are learned by first considering all samples as equally important, then, for the training of the second iteration of the weak classifier, the weighting of all misclassified samples is increased by adjusting the weights of the feature vectors. Hence, the second classifier will focus especially on the previously misclassified samples and therefore on the features that better separate the remaining data. Such strategy has proven to work well for handcrafted features arising from orthogonal and complementary measurements~\cite{Schapire}.

In this work, we used the multi-class extension of Adaboost presented in~\cite{AdaBoost} which performs a forward stagewise additive modeling using a multi-class exponential loss function. We corroborated through a quantitative evaluation strategy (not shown) that this method overperforms other state-of-the-art machine learning models such as SVM's and densely connected multilayer perceptron Neural Networks. The predictions were performed using as a feature set the ensemble of both SMT-dMRI and MET2 derived microstructure parameters:
$F_{comb}$ = $\{SMT_{f_{I}}$, $SMT_{f_{E}}$, $MET2_{f_{M}}$, $MET2_{f_{IE}}$, $MET2_{f_{CSF}}\}$. For the evaluation, we maintain a MS patient dataset which was never used for training the data in any of the repetitions. Three different experiments were performed. In each of these three settings, the classifier was trained and every repetition is used to predict to which category the voxels in the validation image belong.  The classifier implementation was performed using the Scikit-learn 0.22.1 weight-Boosting module. The trained models have the next structure: \emph{Experiment-1}: voxel classification in three different classes (L), (N), (C). \emph{Experiment-2}: lesions versus normal appearing WM, i.e., (L) vs (N). \emph{Experiment-3}: lesions versus healthy controls, i.e., (L) vs (C).  For each experiment, we performed k-fold cross-validation with $k=5$. The feature importance was also computed based on the information gain, which correlates to the feature importance to determine the splits over the aggregation procedure. Finally, instead of reporting a discrete label we report the classification probability between lesions (L) and not lesions (N) or (C) as our "probability estimate map".
\vspace{-3.5mm}
\section{Results}
\vspace{-3.5mm}

\begin{figure}[htp]
    \centering
    \includegraphics[width=8.5cm]{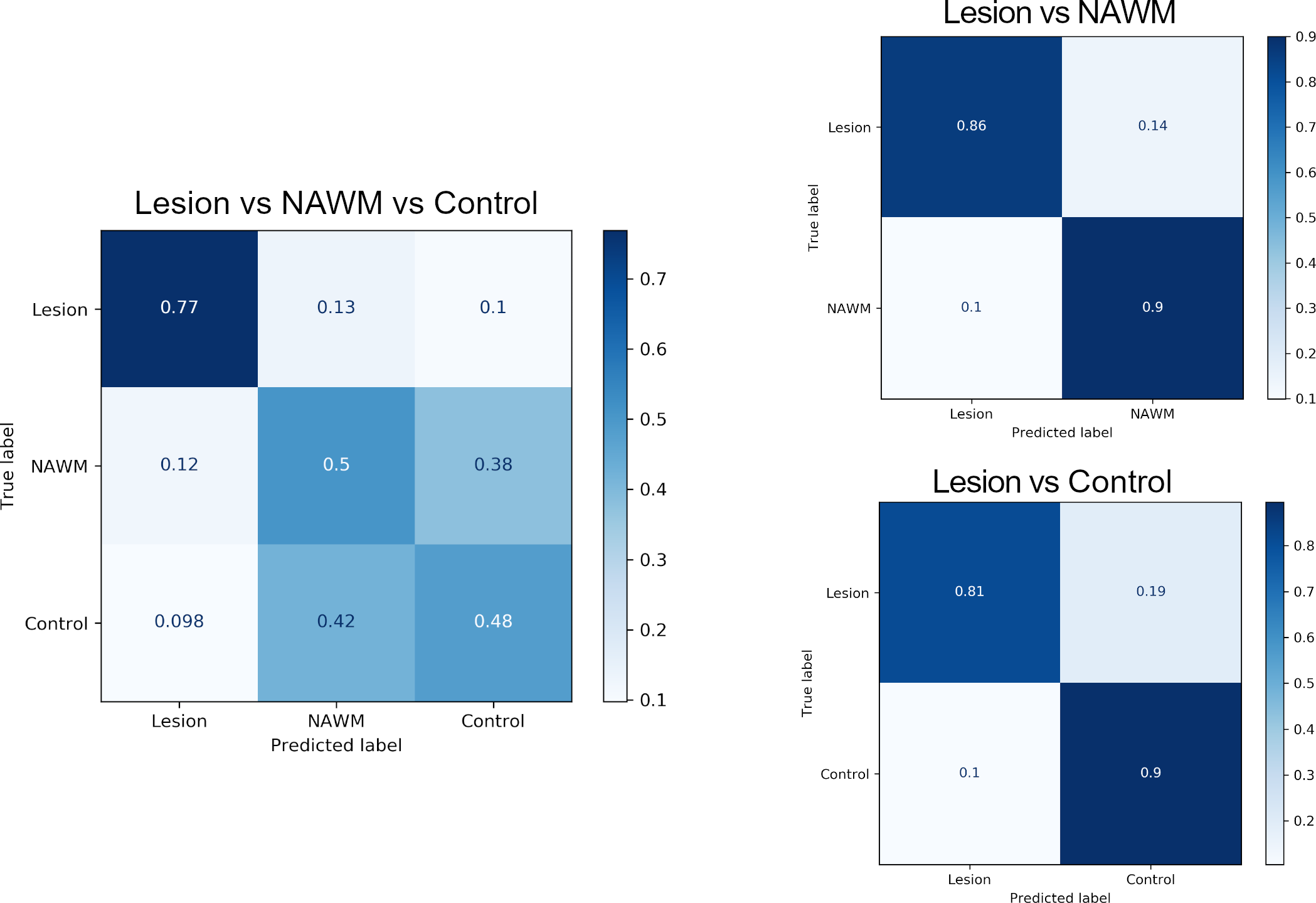}
    \caption{Confusion matrix of the three trained models. The y-axis shows the true label vs the predicted label in the x-axis.}
    \label{fig:conf_matrix}
\end{figure}

Results from the three experiments are summarized in Fig.~\ref{fig:conf_matrix} and Table \ref{tab:test_crossval}, which show, respectively, the confusion matrices of the classification and the cross-validation scores and mean accuracy for the three trained models. Table \ref{tab:feature_importance} shows the mean accuracy of each model trained separately from the diffusion based features ($SMT_{f_{I}}$, $SMT_{f_{E}}$) and the Multi-component T2 features ($MET2_{f_{M}}$, $MET2_{f_{IE}}$, $MET2_{f_{CSF}}$).

\begin{table}[!h]
\centering
\footnotesize
\captionsetup{font=small}
\begin{tabular}{|l|lllll|l|}
\hline
Test scores   & k-1   & k-2   & k-3   & k-4   & k-5   & \begin{tabular}[c]{@{}l@{}}mean\\ accuracy\end{tabular} \\ \hline
\textbf{L-N-C} & 0.598 & 0.607 & 0.600 & 0.597 & 0.593 & \textbf{0.599}                                          \\ \hline
\textbf{L-N}   & 0.843 & 0.861 & 0.851 & 0.841 & 0.840 & \textbf{0.847}                                          \\ \hline
\textbf{L-C}   & 0.858 & 0.853 & 0.848 & 0.843 & 0.854 & \textbf{0.851}                                          \\ \hline
\end{tabular}
\caption{5-folds cross-validation scores and mean accuracy for the three trained models on the test dataset.}
\label{tab:test_crossval}
\end{table}

\begin{table}[!h]
\footnotesize
\captionsetup{font=small}
\begin{tabular}{|l|l|l|l|}
\hline
Mean accuracy & Diff\textsubscript{feat} & MTE\textsubscript{feat} & Diff \& MET2 \textsubscript{feats} \\ \hline
L-C-N    & 0.505            & 0.528           & \textbf{0.599}             \\ \hline
L-N      & 0.785            & 0.703           &  \textbf{0.847}         \\ \hline
L-C      & 0.651            & 0.669            & \textbf{0.851}
                                 \\ \hline
\end{tabular}
\caption{Mean cross-validation accuracy of the classifier for  the diffusion based features (Diff\textsubscript{feat}) and the Multi-component T2 features (MTE\textsubscript{feat}).}
\label{tab:feature_importance}
\end{table}

\begin{figure}[htp]
    \centering
    \includegraphics[width=8.5cm]{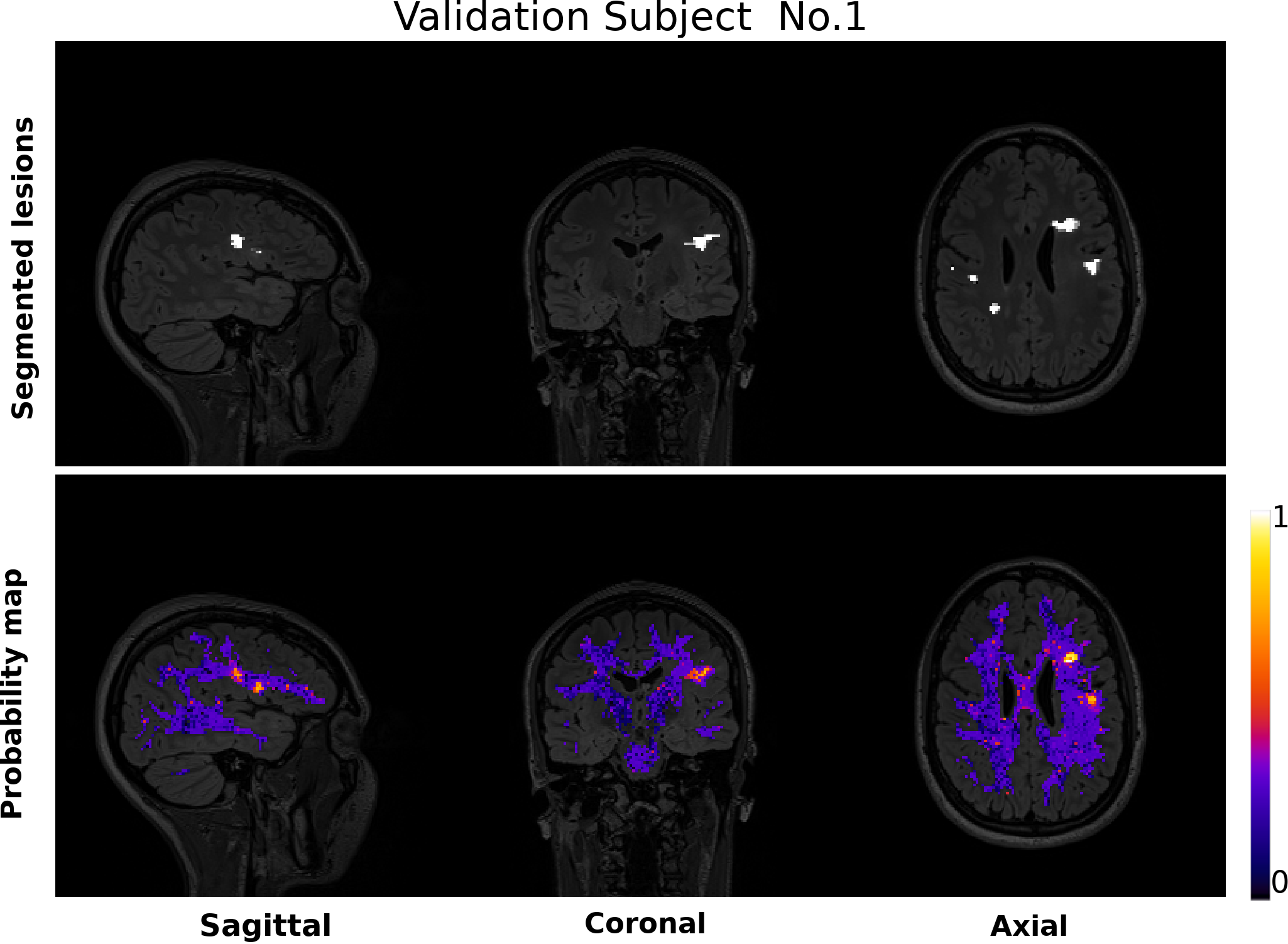}
    \includegraphics[width=8.5cm]{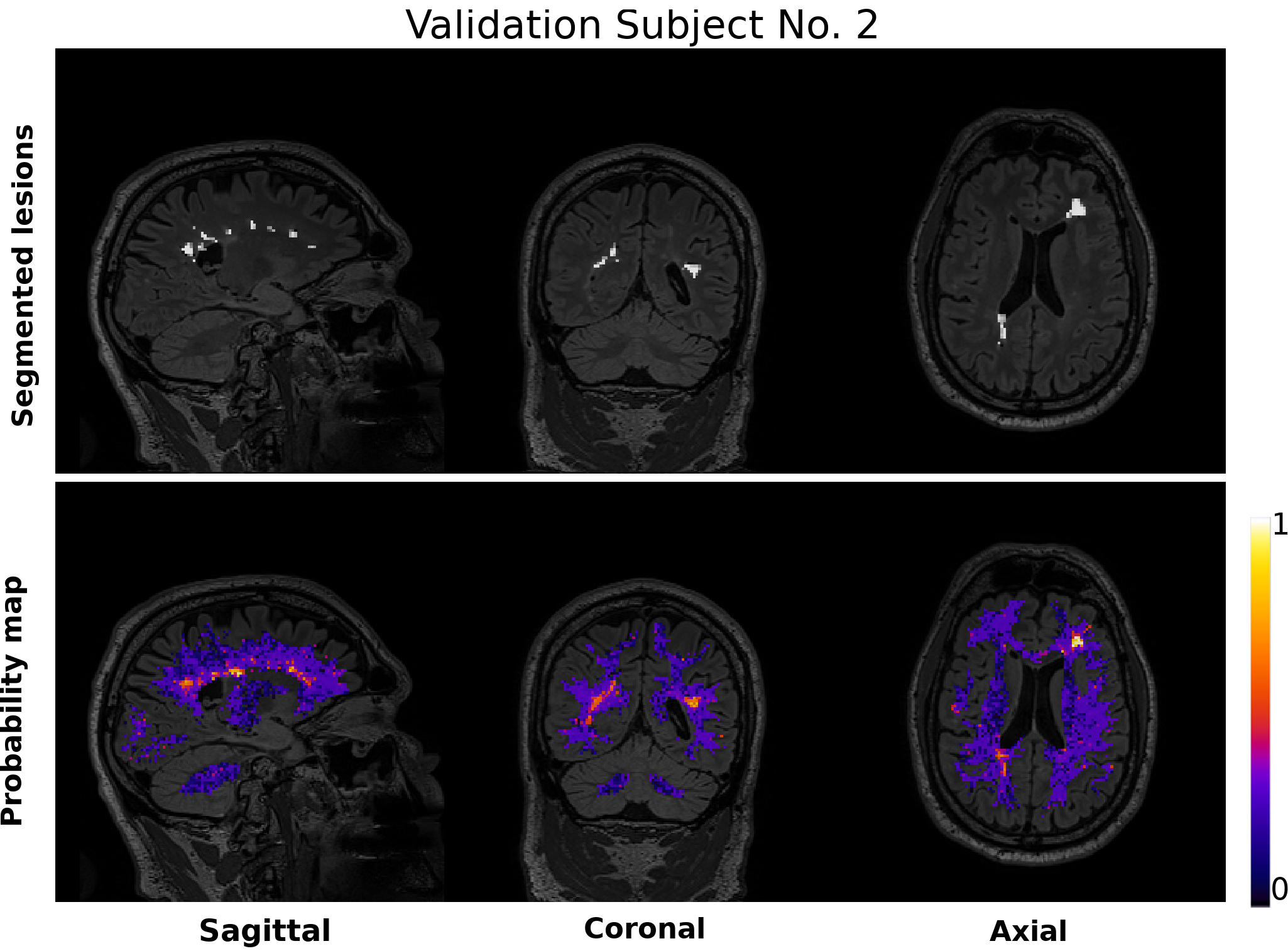}
    \caption{Voxel-wise lesion probability maps for two MS subjects used in the validation. In each panel, the first row corresponds to the sagittal, coronal and axial views over the FLAIR image of the subject's lesions mask. In the second row, same views of the lesion probability map as computed using our proposed method. The probability maps show a gradient in the lesions, with the core of the lesions having a higher lesion score, and a diffuse pattern of abnormal tissue in regions otherwise considered as non-lesional.}
    \label{fig:patients}
\end{figure}

Results for the first setting (L-N-C classification) show that the combination of all features lead to a lesion detection of $77\%$ (Fig.~\ref{fig:conf_matrix} first panel). Most interestingly, the rate of false positives (i.e when a healthy tissue voxel (C) is classified as lesion (L)) is less than $1\%$ (Fig.~\ref{fig:conf_matrix}, first column). As expected, the classification of between NAWM and control voxels (N) and (C) yielded to a modest classification, with almost the same percentage of true and false positives.
Fig.~\ref{fig:patients} compares the lesion probability map for two exemplary RRMS subjects extracted from the testing dataset using our proposed method with the lesion segmentation masks extracted using [15]. While the lesions were correctly classified by our method, the resulting map highlighted voxels within the non-lesional tissue showing a diffuse abnormal pattern. These voxels were mainly located in the perilesional voxels. Moreover, voxels within the MS lesions show a gradient pattern, with the center of the lesion having a higher score than the peripherial voxels.

\begin{figure}[htp]
    \centering
\includegraphics[width=8.5cm]{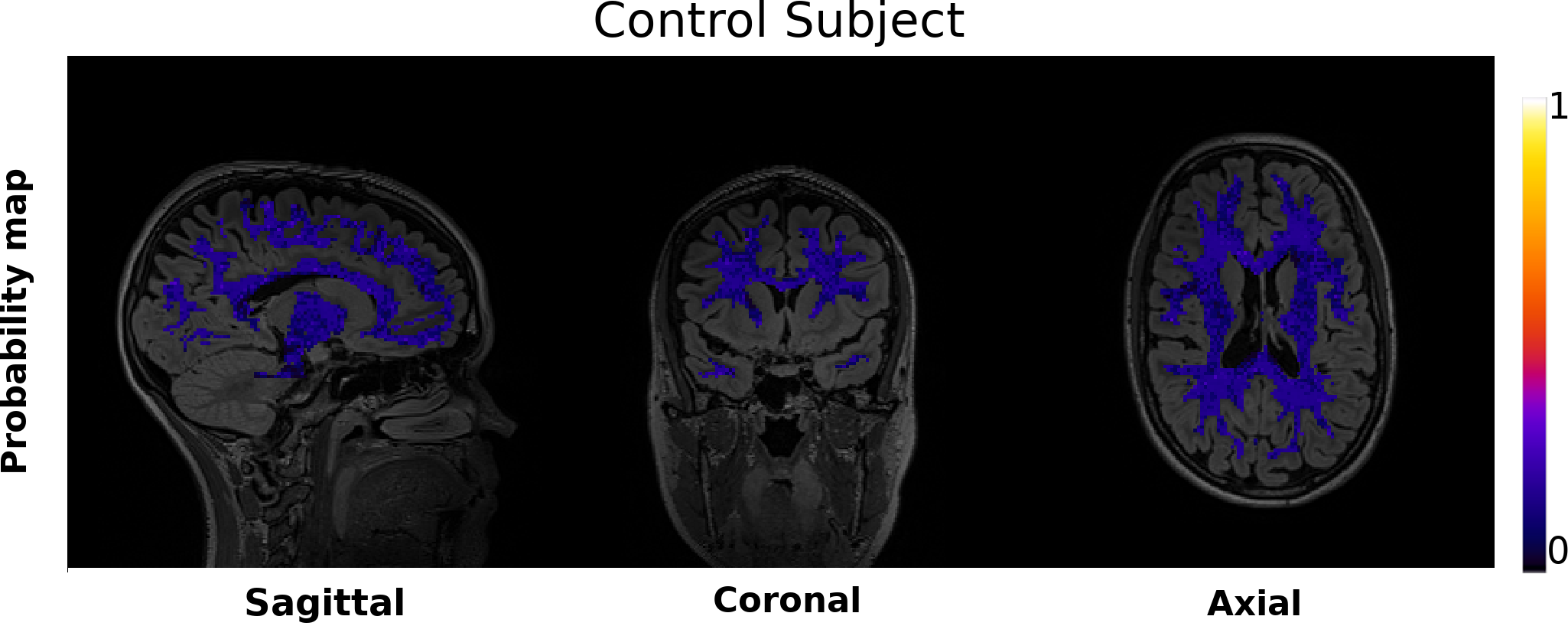}
    \caption{Voxel-wise lesion probability maps for one control subjects used in the validation. Notably, all the WM voxels of the control subject where classified with an overall lower probability than that of the RRMS patients.}
    \label{fig:control}
\end{figure}

\vspace{-3.5mm}

\section{Discussion and Conclusions}
\vspace{-3.5mm}

Several basic processes drive the formation and evolution of MS lesions, like inflammation, myelin breakdown, astrogliosis, oligodendrocyte injury, neurodegeneration, axonal loss, and remyelination. These pathological processes lead to variable and complex changes in different quantitative MRI measures. Our analysis shows that combining tissue microstructure information from multi-compartment dMRI and T2 relaxometry helps at yielding better delineation of MS lesions, compared to the state-of-the-art method used for the segmentation. More interestingly, the lesion probability map as computed using our Ada-Boost classifier, highlighted a gradient pattern in the lesional voxel, in concordance with previous studies showing a progressive tissue loss in the core of MS lesions \cite{Klistorner2018}.
In contrast to recent studies in MS using the fractional anisotropy metric derived from DTI \cite{Ilona2018,Klistorner2018,Chatterjee2018}, in our study we employed a multi-shell acquisition protocol and a more complex biophysical model of the dMRI signal. The parameters estimated for this multi-compartment microscopic diffusion model are unconfounded by fiber crossings and orientation dispersion, thus providing more specific biomarkers of the tissue microstructure. Our study has certain limitations, mainly arising from the limited dataset available. Further work will include increasing the training and testing dataset as well as investigating other dMRI and MET2-based features.

In conclusion, in this work we propose to combine multi-component diffusion MRI and T2 relaxometry models to characterize brain lesions in patients with multiple sclerosis. Our results show that the combination of the dMRI and T2 relaxometry features, together with an Ada-boost classifier add valuable complementary information for the classification of abnormal tissue otherwise considered as non-lesional. The proposed approach is fully automatic and may helps to monitor illness progression, and to assess the efficacy of new treatments.

\vspace{-5.5mm}
\section{Acknowledgments}
\label{sec:acknowledgments}
\vspace{-2mm}
\small{This work is supported by the Strategic Focal Area “Personalized Healthcare and Related Technologies” of the ETH domain grant 2018-425 to EFG, the European Union’s Horizon 2020 under the Marie Skłodowska-Curie grant 754462 to MP, the Swiss National Science Foundation grant PZ00P2\_185814/1 to EJC-R and the Centre for Biomedical Imaging of the University of Lausanne, the Swiss Federal Institute of Technology Lausanne, the Lausanne University Hospital to J-PT. GFP, TH and TK work for Siemens Healthineers AG, Switzerland.}

\bibliographystyle{IEEEbib}

\end{document}